\pdfoutput=1

\documentclass[11pt]{article}

\usepackage{acl}

\usepackage{times}
\usepackage{latexsym}

\usepackage[T1]{fontenc}

\usepackage[utf8]{inputenc}

\usepackage{microtype}

\usepackage{inconsolata}

\usepackage{nicematrix}
\usepackage{caption}
\usepackage{subcaption}
\usepackage{placeins}
\usepackage{enumitem}
\usepackage{hyperref}
\usepackage{tcolorbox}
\usepackage{listings}
\usepackage{multirow}
\tcbuselibrary{listings, skins} 
\title{Extractive Schema Linking for Text-to-SQL}

\author{
    Michael Glass\textsuperscript{1},
    Mustafa Eyceoz\textsuperscript{2},
    Shankar Subramanian\textsuperscript{1}, \\
    {\bf Gaetano Rossiello\textsuperscript{3},
    Long Vu\textsuperscript{1},
    Alfio Gliozzo\textsuperscript{1}} \\
    \\
    IBM Research AI \\
    \textsuperscript{1} \texttt{\{mrglass, dharmash, lhvu, gliozzo\}@us.ibm.com} \\
    \textsuperscript{2} \texttt{mustafa.eyceoz@partner.ibm.com} \\
    \textsuperscript{3} \texttt{gaetano.rossiello@ibm.com}
}

\begin{document}
\maketitle
\begin{abstract}

Text-to-SQL is emerging as a practical interface for real world databases. The dominant paradigm for Text-to-SQL is cross-database or schema-independent, supporting application schemas unseen during training.
The \textit{schema} of a database defines the tables, columns, column types and foreign key connections between tables.
Real world schemas can be large, containing hundreds of columns,  but for any particular query only a small fraction will be relevant.  Placing the entire schema in the prompt for an LLM can be impossible for models with smaller token windows and expensive even when the context window is large enough to allow it.
Even apart from computational considerations, the accuracy of the model can be improved by focusing the SQL generation on only the relevant portion of the database. \textit{Schema linking} identifies the portion of the database schema useful for the question.

Previous work on schema linking has used graph neural networks, generative LLMs, and cross encoder classifiers.
We introduce a new approach to adapt decoder-only LLMs to schema linking that is both computationally more efficient and more accurate than the generative approach.  Additionally our extractive approach permits fine-grained control over the precision-recall trade-off for schema linking.
\end{abstract}

\section{Introduction}

Databases play a crucial role in the realm of business and various other domains due to the wealth of information they contain. However, harnessing this abundance of data to effectively address important queries and gather valuable insights can often prove to be challenging.
Text-to-SQL is an approach to constructing queries over databases by using natural language which is then translated to Structured Query Language (SQL).  Because Text-to-SQL offers the potential to enable anyone to gather insights, generate reports, answer questions and create dashboards backed by live data it has received a lot of attention from the research community 
\cite{rai-etal-2023-improving}, \cite{c3}.

Databases are structured according to a schema, providing metadata on the set of tables and the columns within each table. Each column is defined by its name and datatype, and the schema may include constraints such as primary keys and foreign keys. Figure \ref{fig:schema_link_input} illustrates a simple example of a database schema using Data Definition Language (DDL) statements. These `CREATE TABLE' statements represent one method of defining a database schema for a Text-to-SQL system \cite{dail-sql}.

\begin{figure*}
    \centering
    \includegraphics[width=0.9\linewidth]{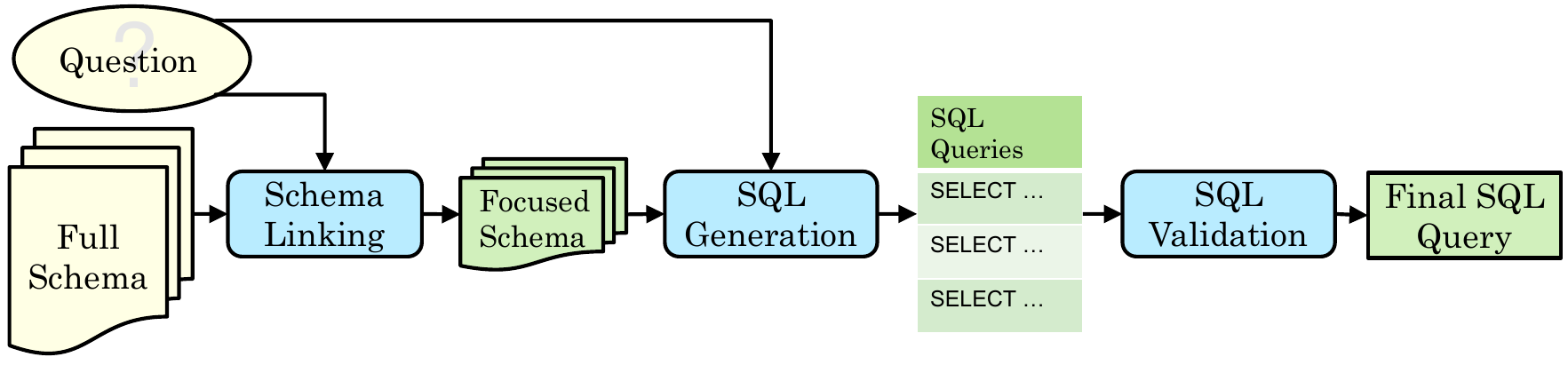}
    \caption{Text-to-SQL System Architecture Overview}
    \label{fig:Text-to-SQL_system}
\end{figure*}

Early work on Text-to-SQL \cite{geoquery}, \cite{iyer-etal-2017-learning} focused on training a model for a specific database. More recently, Text-to-SQL systems are trained to be schema-independent \cite{wikisql}, \cite{yu-etal-2018-spider}.  A schema-independent system can generate SQL queries from natural language questions over any provided database schema, rather than being trained to construct queries over only one database.  This is important for two reasons: first, we can gather question-SQL pairs over training schemas to train the system and then deploy to unseen business schemas; second, the schema can change, growing and adapting to requirements while the Text-to-SQL system continues to function.  Because a schema-independent Text-to-SQL system is expected to work across schemas, the schema must be provided along with the natural language question as an input.  However, this introduces another problem: the schema can be very large while only a small fraction will be needed to generate the query.

The size of the schema presents several challenges: token limits, computational costs, and the need for precision in SQL generation. Modern Text-to-SQL systems rely on foundational Large Language Models (LLM), particularly those trained extensively on code \cite{scholak-etal-2021-picard}, \cite{shaw-etal-2021-compositional}, \cite{dail-sql}. While recent LLMs permit large token counts, smaller, more efficient LLMs have lower token limits, making it challenging to accommodate an entire schema. Additionally, even if the token limit is not surpassed, each input token contributes to increased computational expense. Beyond computational benefits, providing only the pertinent parts of the schema to the LLM during SQL generation significantly reduces the likelihood of producing erroneous queries. A focused input schema can further enhance query accuracy by allowing for the inclusion of richer details in the schema description, such as example rows for each relevant table. 

Figure \ref{fig:Text-to-SQL_system} is a basic three phase Text-to-SQL system e.g. \citet{resdsql} or \citet{din-sql}.  The question and the full schema are the input to schema linking, which identifies a subset of the schema that is relevant.  SQL generation uses this focused schema and the question to generate candidate SQL queries.  SQL validation selects among these candidates to identify the final predicted SQL query.  The most basic validation simply rejects SQL statements that can not execute \cite{suhr-etal-2020-exploring}.  

Our contributions are: 1) a new model for schema linking producing a probability prediction over the hidden states of a decoder-only LLM, 2) fine-grained prediction of the role each column plays in the SQL query, 3) an extensive evaluation establishing a new state-of-the-art in schema linking, 4) a controllable trade-off between precision and recall and an empirical examination of their relative importance for impact in SQL generation.

\section{Related Work}

RAT-SQL \cite{wang-etal-2020-rat} represents an early approach to Text-to-SQL using a pre-LLM schema linking methodology. It combines GloVe embeddings \cite{pennington-etal-2014-glove} with a BiLSTM \cite{lstm} to process question, column, and table names, generating vector representations for schema elements and query terms. 
Subsequently, a Graph Neural Network (GNN) models the relationships between tables and columns, informed by the database schema. Relationships between question tokens and schema elements are established through n-gram matching for names and value matching for column contents.

In the realm of LLM-based schema linking, there are primarily two approaches: generative and cross-encoder. A generative approach is a natural choice since the SQL generation uses a generative LLM, and can be implemented in either a few-shot or a fine-tuned manner.

DIN-SQL \cite{din-sql} utilizes GPT-4 \cite{gpt-4} with decomposed in-context learning for schema linking, SQL generation, and SQL debugging. This process is facilitated by prompt engineering tailored for GPT-4, where few-shot prompts are enhanced with a chain-of-thought technique. For instance, schema linking prompts begin with the directive ``Let's think step by step,'' followed by the question. In the few-shot examples, this is followed with a step where the question is repeated and the value or values requested are connected to columns of specific tables. Subsequent steps identify necessary joins and specify cell values needed in the WHERE clause, culminating in a summary of all schema links for straightforward extraction from the generated text. While intuitive, this approach has limitations in performance and lacks a mechanism to provide confidence scores for the schema links.

DTS-SQL \cite{dts-sql} adopts a generative approach for schema linking at the table level, predicting a list of relevant tables based on the given question and the entire database schema. This method involves fine-tuning models for both schema linking and SQL generation.

Another prominent method involves the use of a transformer encoder, or cross-encoder. RESDSQL \cite{resdsql} pairs the natural language question with the database schema as inputs to a cross-encoder, utilizing the encoder-only model of RoBERTa \cite{roberta} to generate vector representations of schema elements. These vectors are pooled using a BiLSTM, with an additional attention layer enhancing the focus on relevant tables and columns. This information is then used to classify schema elements through a fully connected layer. While effective, this approach is limited to encoder-only or encoder-decoder models, contrasting with the decoder-only architecture common to many contemporary code LLMs, such as StarCoder \cite{starcoder}, WizardCoder \cite{wizardcoder}, and DeepSeek Coder \cite{deepseek-coder}.

\section{Approach}

We combine the key strength of the generative approach: using large, modern code LLMs, with the strength of the cross-encoder approach: a probability for each schema item, enabling recall-oriented predictions.  Furthermore, we introduce \textit{fine-grained} schema linking which predicts a probability for each role a database column could play in the SQL query.

\begin{figure}[thb]
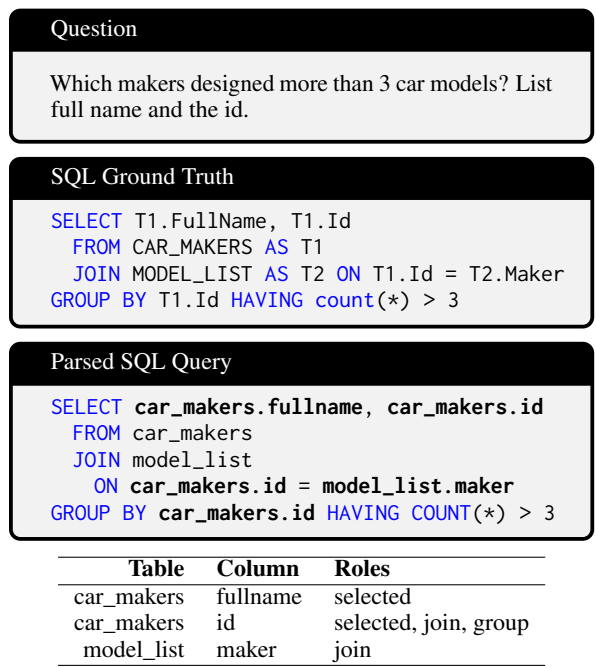

    \centering
    \small
    \begin{tcolorbox}[colback=gray!10,colframe=black,title=Question]
    Which makers designed more than 3 car models? List full name and the id.
    \end{tcolorbox}
    
    \begin{tcolorbox}[colback=gray!10,colframe=black,title=SQL Ground Truth]
    \begin{lstlisting}[basicstyle=\ttfamily,columns=fullflexible,aboveskip=-5pt,belowskip=-5pt]
SELECT T1.FullName, T1.Id 
  FROM CAR_MAKERS AS T1 
  JOIN MODEL_LIST AS T2 ON T1.Id = T2.Maker 
GROUP BY T1.Id HAVING count(*) > 3
    \end{lstlisting}
    \end{tcolorbox}

    \begin{tcolorbox}[colback=gray!10,colframe=black,title=Parsed SQL Query]
    \begin{lstlisting}[basicstyle=\ttfamily,columns=fullflexible,aboveskip=-5pt,belowskip=-5pt]
SELECT @car_makers.fullname@, @car_makers.id@
  FROM car_makers 
  JOIN model_list 
    ON @car_makers.id@ = @model_list.maker@ 
GROUP BY @car_makers.id@ HAVING COUNT(*) > 3
    \end{lstlisting}
    \end{tcolorbox}
    
    \begin{tabular}{rll}
    \hline
    \textbf{Table} & \textbf{Column} & \textbf{Roles} \\ 
    \hline
    car\_makers & fullname & selected\\ 
    car\_makers & id & selected, join, group \\
    model\_list & maker & join\\
    \hline
    \end{tabular}

\caption{Ground truth generation for schema linking}
     \label{fig:ground_truth}
\end{figure}

\subsection{Data Preparation}
The first step in training a schema linker is to gather ground truth.  
A Text-to-SQL dataset can also provide ground truth for schema linking by examining the SQL queries to find what tables and columns are used in the query. 
Earlier approaches used a simple token matching strategy to find the tables and columns referenced by a SQL statement~\cite{resdsql}. This is accurate enough to train an effective model, but it is not usable for an evaluation of the schema linker.  We found that a majority of mismatches between our schema linker and the ground truth were problems with the ground truth.

Figure \ref{fig:ground_truth} illustrates our approach. Rather than token matching, we parse and analyze the SQL statements for each question, simplifying table aliases as needed to identify the columns of each table that are referenced. We use mo-sql-parsing\footnote{\url{https://github.com/klahnakoski/mo-sql-parsing}} for this static analysis.
This produces a ground truth where each natural language question, plus the associated database schema is mapped to a set of relevant, qualified column names, i.e. \texttt{table.column}.  

When parsing we also build an indicator vector for each column, indicating what roles it plays in the SQL query.
By predicting the roles for each column we assist the downstream SQL generation in constructing the query. The roles we consider are: 

\begin{itemize}[noitemsep]
    \item \textit{selected}: used in any SELECT, either outermost or a sub-select.
    \item \textit{join}: used in any JOIN ON statement.
    \item \textit{condition}: involved in any comparison other than joins.
    \item \textit{order}: used in an ORDER BY clause.
    \item \textit{group}: used in a GROUP BY clause.
\end{itemize}

Sometimes a table is relevant for a question, but no specific column is needed. For example, in a query such as \lstinline|SELECT COUNT(*) FROM singer|, the \texttt{singer} table is needed but no specific column is needed. In this case, we simply indicate the first column from the table is relevant.

In contrast, SL-SQL~\cite{slsql} uses a mix of automatic and manual annotation to produce a ground truth for schema linking that also connects the tables and columns in the relevant schema to specific words in the question. We compare our fully automatically generated ground truth to this annotation and find many differences. Out of 7000 Spider training examples there are differences in 4816.  Most of these cases are columns that are not mentioned explicitly in the question but are necessary for joining two tables.  Considering only those cases where the SL-SQL annotation contains additional columns or tables not found by our parsing of the ground truth SQL, we find 168 cases.  Most of these involve a table or column that is a close string match for some word in the question, but is not needed for the query.  Rarely the SL-SQL annotation is correct while the ground truth SQL query is incorrect. Appendix \ref{sec:sl-sql-gt} shows examples of these cases.

We use our approach to gather ground truth for both training our schema linkers and evaluating them at the component level.

\subsection{Generative Schema Linking}\label{gensl}
As an initial experiment we implemented a fine-tuned generative approach to schema linking.  In this approach we train a generative model to produce a comma separated list of qualified column names.  At inference time, many samples are drawn from the generative model to produce multiple candidate lists of relevant columns.  This both increases recall and provides a crude confidence estimation by scoring each column according to how many times it is generated.  However, fine-tuning open source code LLMs for generative schema linking did not match the accuracy of state-of-the-art systems based on GPT-4~\cite{gpt-4}. 

\subsection{Extractive Schema Linking}\label{exsl}
To address the deficits of the generative approach, we developed a decoder-only extractive approach to schema linking: \textit{ExSL}.  In the extractive approach, rather than generate new tokens autoregressively, we use the final vector representation of key input tokens to predict which parts of the database schema are relevant.  
This is similar to the cross-encoder model of RESDSQL \cite{resdsql}.  
However, with the exception of CodeT5 \cite{wang-etal-2021-codet5}, modern code LLMs are decoder-only models.  Therefore to make use of the code knowledge these models obtain in pre-training, we adapt the extractive approach to a decoder-only model. 


Note that in a decoder-only model, the final hidden vector for each token only considers itself and the tokens before it \cite{attention}.  Since it is likely beneficial to see the entire schema before determining what is relevant, we repeat the columns as a list of candidates after the entire schema and the question.  In this way, predictions for relevance are always made with the entire question and schema within the attention mask.  

\begin{figure}[h!]
\small
    \centering
    \begin{tcolorbox}[boxsep=0mm, boxrule=0.5pt, colback=white] 
        \begin{lstlisting}[basicstyle=\ttfamily,columns=fullflexible,aboveskip=-4pt]
CREATE TABLE publication (
  publication_id NUMBER PRIMARY KEY,
  book_id NUMBER,
  publisher TEXT,
  publication\_date TEXT,
  price NUMBER,
  FOREIGN KEY(book_id)
     REFERENCES book(book_id) );

CREATE TABLE book (
  book_id NUMBER PRIMARY KEY,
  title TEXT,
  issues NUMBER,
  writer TEXT );
\end{lstlisting}
To answer:
Show the titles of books in descending order of publication price.\\
We need columns:\\
« publication publication\_id»\\
« publication book\_id»\\
« publication publisher»\\
« publication publication\_date»\\
« publication price»\\
« book book\_id»\\
« book title»\\
« book issues»\\
« book writer»
\end{tcolorbox}
    \caption{Example schema linker input}
    \label{fig:schema_link_input}
\end{figure}

\begin{figure}[h!]
    \centering
\small
    \begin{tcolorbox}[boxsep=0mm, boxrule=0.5pt, colback=white] 
        \begin{lstlisting}[basicstyle=\ttfamily,columns=fullflexible,aboveskip=-4pt]
CREATE TABLE publication (
    book_id NUMBER,
    price NUMBER,
    FOREIGN KEY(book_id)
        REFERENCES book(book_id) );

CREATE TABLE book (
    book_id NUMBER PRIMARY KEY,
    title TEXT );
        \end{lstlisting}
        \begin{tcolorbox}[colback=gray!10] 
            selected: book.title \\
            join: book.book\_id, publication.book\_id \\
            condition: None \\
            order: publication.price \\
            group: None
        \end{tcolorbox}
        
        Show the titles of books in descending order of publication price.
    \end{tcolorbox}
    \caption{Focused schema input}
    \label{fig:focused_schema_input}
\end{figure}

If the entire schema is too large, the schema linking task is split across multiple instances by greedily placing as many tables as will fit in the token window into each instance.  The columns for each subset of the tables will have their relevance probabilities predicted independently. The relevance probabilities for the full schema is then simply the union of all the subsets. 

Figure \ref{fig:schema_link_input} shows a diagram of this input representation. 
For each pair of candidate column marks (`«' and `»'), the final token embeddings, $E_{\alpha_i}$ and $E_{\omega_i}$,  are concatenated to produce a single vector. Then a linear layer, $\mathbf{w_{relevance}}$, is applied to predict the likelihood for each candidate that it is relevant for the question.  In the case of fine-grained schema linking, this predicts a probability for each possible role, possibly multiple roles per-column if the column is used in multiple clauses. 

\begin{align*}
\alpha & = [i : t_{i} = ``\text{«}" ] \\
\omega & = [i : t_{i} = ``\text{»}" ] \\
C & = \begin{bmatrix}
E_{\alpha_0} \oplus E_{\omega_0} \\
E_{\alpha_1} \oplus E_{\omega_1} \\
E_{\alpha_2} \oplus E_{\omega_2} \\
...
\end{bmatrix} \\
\rho & = C \cdot \mathbf{w_{relevance}}
\end{align*}

To train the model we use a binary cross entropy loss against the relevance labels obtained by parsing the ground truth SQL queries for all mentioned columns.   
Details of the hyperparameters used are in the appendix.

\subsection{Focused Schema}

After applying the extractive schema linker there is a probability prediction for each qualified column.  In order to determine what subset of the schema is deemed relevant, and therefore included in the SQL generation prompt, we need to set a threshold or select a top-k.  Since there is considerable variation in the number of needed columns for different queries, a threshold is more appealing than selecting the top-k.  Intuitively, recall is more important than precision when selecting a threshold.  The SQL generation can ignore columns that are not needed for the query.  However, without a needed column in recall, the SQL generation will be unable to generate the correct query, unless it is lucky enough to guess the name of the needed column.  In section \ref{sec:pr_tradeoff} we explore the impact of selecting the threshold and validate our intuition that recall is more important than precision.

Figure \ref{fig:focused_schema_input} shows a focused schema. After schema linking, only the columns with predicted relevance probability above the threshold are retained as input to the SQL generation. The gray box shows the format for our fine-grained schema linking output in the prompt for SQL generation.  We provide a list of columns for each role sorted by their predicted probability for that role. When no columns are relevant for a role (logit threshold -3.0) `None' is provided.

The schema representation can be improved by providing sample data for each of the relevant tables, considering only the relevant columns.  ACT-SQL \cite{act-sql} finds this to be an effective way to elaborate on the schema description.  Without schema linking this enrichment is impractical for large schemas and empirically counterproductive even for the modest schema sizes in Spider.  



\subsection{SQL Generation}
To assess the end-to-end impact of the schema linking we also train SQL generation models. This model takes the subset of the schema selected by the schema linker along with the natural language question and generates the SQL query.  We train separate models to consume the column-level, ExSL$_c$, and the fine-grained, ExSL$_f$, focused schemas.

The schema linking model produces a query-pertinent subset of the database schema to feed a downstream SQL generation model. For overall end-to-end performance, the SQL generation model needs to be robust to noise from the upstream schema linking. We present a strategy to prepare the training data for a SQL generation model toward such end-to-end robustness. 
The training data to fine-tune an LLM for SQL generation could use either the full schema or the exact subset of the schema required for the SQL query in the input prompt, with the corresponding ground truth SQL query as the target.
Using the full schema is either impossible in the case of very large schemas due to input token length limits, or expensive. Training on the exact ground truth schema subset leaves the SQL generation model unprepared for inevitable errors from the schema linker. 

We propose to make the SQL generation model robust toward flawed schema linking with a more careful choice of the training data set. 
For each training data instance in Text-to-SQL training data sets, we first construct the set difference between the full schema set and the exact query-pertinent schema subset. We then sample randomly from this set difference to introduce a controlled level of ``noise''. For example, if the set difference contains 10 elements, 20\% noise would correspond to any two elements sampled from this set and added to the ground truth schema subset. 
This exposes the SQL generation model to diverse, noisy inputs for the focused schema, thereby imparting robustness by mimicking errors from the upstream schema linker. 

When applying the SQL generation we generate with beam search.  Each generated query is then executed against the database.  Any query that fails to execute will be discarded and the next query in the beam is attempted instead. With this strategy nearly all questions have at least one executable query.

\section{Datasets}\label{sec:experiments}

\begin{table*}[!htb]
    \centering
    \begin{minipage}[t]{0.31\linewidth}
        \centering
        \begin{NiceTabular}{l|rr}
        \hline
        Split & Instances & Schemas \\
        \hline
        Train     & 7000 & 140 \\
        Dev       & 1034 & 20 \\
        Test      & 2147 & 40 \\
        DK        & 535  & 10 \\
        Syn       & 1034 & 20 \\
        Real & 508  & 19 \\
        \hline
        \end{NiceTabular}
        \caption{Statistics for Spider and its variants}
        \label{tab:datasets}
    \end{minipage}\hfill
    \begin{minipage}[t]{0.67\linewidth}
        \centering
        \begin{tabular}{lcccccc}
        \hline
        \multirow{2}{*}{Method} & \multicolumn{2}{c}{BIRD-dev} & \multicolumn{2}{c}{Spider-dev} & \multicolumn{2}{c}{Spider-test} \\
        \cline{2-7}
        & P & R & P & R & P & R \\
        \hline
        DTS-SQL  & 95.07 & 92.74 & 98.48 & 97.77 & 97.33 & 98.16 \\
        Gen      & 90.40 & 95.50 & 95.16 & 99.32 & 95.70 & 98.96 \\
        ExSL$_{c}$ & 95.86 & 93.94 & 98.87 & 98.80 & 98.35 & 98.83 \\
        ExSL$_{f}$ & 96.35 & 93.85 & 98.64 & 98.92 & 98.29 & 98.32 \\
        \hline
        \end{tabular}
        \caption{Schema linking table-level evaluation}
        \label{tab:table_level_eval}
    \end{minipage}
\end{table*}

We train and evaluate on the popular Spider dataset \cite{yu-etal-2018-spider} (CC BY-SA 4.0).  Spider is the most well studied cross-database Text-to-SQL dataset and allows comparison of our schema linker with the previous state-of-the-art. The English questions are categorized as easy, medium, hard and extra-hard based on the required complexity of the SQL query.
We also consider three variants of the Spider dataset, developed to create additional challenges and test the robustness of Text-to-SQL systems: Spider-DK (Domain Knowledge) \cite{spider-dk}, Spider-Syn (Synonyms) \cite{spider-syn}, and Spider-Realistic \cite{spider-realistic}

    
    

Table \ref{tab:datasets} shows the basic statistics for the datasets we used.  In all cases the schemas for training do not overlap with the schemas for dev or test.  

We also evaluate on the recently introduced BIRD dataset \cite{li2024can} (CC BY-SA 4.0), which contains over 12,751 unique question-SQL pairs, covering 95 large databases that span 37 professional fields, including blockchain, healthcare, and education among others. BIRD emphasizes the challenges of noisy database values as well as external knowledge that connects the natural language question to database values.

\section{Evaluation}

Our primary experiments consider the evaluation of the schema linker component at predicting the columns involved in the ground truth SQL statement for each question.   
For comparison to prior art we also consider table-level metrics measuring the precision and recall of tables found by the schema linker.  DTS-SQL uses this metric since it does not produce column-level relevance predictions.

The value of schema linking is justified primarily through its impact in final Text-to-SQL performance.  In our end-to-end evaluation, we compare the accuracy of SQL generation when provided different schema links. We assess that a generated SQL query is correct if it produces the same results as the gold standard SQL query.  We use the Spider \footnote{\url{https://github.com/taoyds/spider}} and BIRD \footnote{\url{https://github.com/AlibabaResearch/DAMO-ConvAI/tree/main/bird}} evaluation scripts to measure execution accuracy.


\subsection{Component Evaluation}

Table \ref{tab:table_level_eval} shows our table-level evaluations. Our baseline generative schema linker produces better recall at the cost of precision. Using either the coarse or fine-grained extractive schema linking  outperforms DTS-SQL simultaneously on both precision and recall across all datasets.  

We compare a simple fine-tuned Generative schema linker, the cross-encoder schema linker of RESDSQL, and the generative chain-of-thought prompting of DIN-SQL (using GPT-4) with our extractive, decoder-only approach.  Both the generative and extractive models are fine-tuned from DeepSeek Coder 6.7B.  

Table \ref{tab:component} has the component level evaluation of the schema linker. 
Unfortunately there is no standard metric for evaluation of schema linking. RESDSQL~\cite{resdsql} uses ROC AUC, while others use precision and recall over different ground truths \cite{slsql}, \cite{eta-awakening}, \cite{lei-etal-2020-examining}.  We report area under the precision / recall curve to summarize precision and recall across different thresholds, the $F_6$ measure to integrate precision and recall with a focus on recall, and ROC AUC for comparison to previous work.

Across all metrics and all datasets our extractive schema linkers outperform both the generative approach and the prior state-of-the-art.  
Coarse-grained schema linking (ExSL$_c$) is typically better for the component metrics of relevant column prediction, but ExSL$_f$ shows impact in Section \ref{sec:e2e}.
Additionally the extractive approach is more than twenty times faster than the autoregressive generative approach. 






\begin{table}[!htb]
\small
\centering
\begin{NiceTabular}{l|rrrrr}
\hline
Linker & Dev & Test & DK & Syn & Real \\
\hline
& \multicolumn{5}{c}{$F_6$ scores} \\
\hline
Gen & 97.15 & 97.68 & 95.07 & 92.59 & 92.25 \\
RESDSQL    & 97.61 & 97.22 & 95.44 & 93.36 & 95.57 \\
DIN-SQL    & 95.32 & - & - & - & - \\
ExSL$_{c}$   & 98.45 & \textbf{98.79} & 97.73 & \textbf{94.36} & \textbf{95.66} \\
ExSL$_{f}$   & \textbf{98.52} & 98.51 & \textbf{98.06} & 93.52 & 94.98 \\
\hline
& \multicolumn{5}{c}{ROC AUC scores} \\
\hline
Gen & 97.00 & 97.45 & 95.39 & 93.63 & 93.42 \\
RESDSQL    & 99.39 & 98.82 & 98.36 & 97.60 & 98.40 \\
DIN-SQL    & 96.54 & - & - & - & - \\
ExSL$_{c}$   & \textbf{99.79} & \textbf{99.74} & 99.43 & \textbf{98.63} & \textbf{98.98} \\
ExSL$_{f}$   & 99.76 & 99.70 & \textbf{99.47} & 98.32 & 98.72 \\
\hline
& \multicolumn{5}{c}{PR AUC scores} \\
\hline
Gen & 78.69 & 84.12 & 73.29 & 67.41 & 73.92 \\
RESDSQL   & 96.51 & 97.04 & 92.42 & 91.31 & 94.14 \\
DIN-SQL    & 92.33 & - & - & - & - \\
ExSL$_{c}$   & \textbf{98.52} & \textbf{98.65} & 96.67 & \textbf{93.54} & \textbf{95.70} \\
ExSL$_{f}$   & 98.44 & 98.40 & \textbf{97.09} & 92.42 & 95.36 \\
\hline
\end{NiceTabular}
\caption{Schema linking column evaluation on Spider}
\label{tab:component}
\end{table}

\begin{table}[ht]
\centering
\begin{tabular}{lccc}
\hline
Linker     & $F_6$ & ROC & PR AUC \\
\hline
Generative & 91.04 & 94.15 & 69.18 \\
ExSL$_c$    & 96.21 & 99.35 & \textbf{93.77} \\
ExSL$_f$    & \textbf{96.45} & \textbf{99.38} & 93.67 \\
\hline
\end{tabular}
\caption{Schema linking column evaluation on BIRD}
\label{tab:linker_performance}
\end{table}

\subsection{End-to-end Evaluation}
\label{sec:e2e}

To assess the value of schema linking on the final SQL generation, we consider three cases: 1) no schema linking, using the full schema, 2) using exactly the relevant schema, mimicking perfect schema linking, 3) using a trained schema linker.  In case three we examine each of the schema linkers considered in our component evaluation. 

Tables \ref{tab:e2e} shows that schema linking with an accurate, recall oriented schema linker is helpful relative to using the entire schema as input.  ExSL$_f$ and ExSL$_c$ also improve substantially over the generative baseline, DIN-SQL, and RESDSQL, reinforcing the conclusion of the component evaluation in Table \ref{tab:component}.  While the component evaluation suggested the fine-grained schema linker, ExSL$_f$, does not improve over ExSL$_c$ for predicting relevant columns, it does show impact in SQL generation by providing additional information. 

Comparing the performance of the trained schema linkers to the ideal schema linker (GT$_{c}$) we see that there is still headroom for more improvement in schema linking.  Interestingly, on DK, ExSL$_{f}$ is even able to outperform the column-level ground truth by providing additional clues about the specific roles of the identified columns to the SQL generation.

Table \ref{tab:e2e_bird} presents our results on the BIRD dev set.  We compare GT$_t$, giving table-level ground truth, GT$_c$, giving column-level ground truth, and GT$_f$, giving fine-grained information on the roles of each column.  We see that increasing the detail of schema linking, if done accurately, provides large impact for SQL generation.  This proves true in practice with ExSL$_f$ improving substantially over the baselines.  Surprisingly, the generative baseline is the most effective for `Challenging' questions.  It does not have a similar advantage for the `Hard' or `Extra hard' questions in Spider, but the most challenging questions in BIRD are significantly harder than the most challenging questions in Spider, so there may be some advantage to predicting schema elements autoregressively for very difficult SQL.  
Due to the larger size of some BIRD schemas, we do not compare to using the full schema.

Appendix \ref{sec:additional_analysis} includes a study correlating schema linking performance to SQL generation accuracy at the instance level. 







\begin{table}[!htb]
\centering
\begin{NiceTabular}{l|rrrrr}
\hline
Schema & Dev & Test & DK & Syn & Real \\
\hline
Full    & 78.7 & 77.3 & 62.4 & 69.7 & 69.5 \\
Gen     & 77.3 & 79.0 & 64.3 & 69.2 & 71.5 \\
\small{DIN-SQL} & 78.9 &      &      &      & \\
\small{RESDSQL} & 80.9 & 80.1 & 69.3 & 74.3 & 75.8 \\
ExSL$_{c}$ & 81.2 & 81.4 & 69.5 & 73.1 & \textbf{77.2} \\
ExSL$_{f}$ & \textbf{82.4} & \textbf{83.0} & \textbf{73.3} & \textbf{74.7} & 75.2 \\
\hline
GT$_{c}$   & 84.0 & 83.5 & 71.2 & 82.3 & 81.7 \\
\hline
\end{NiceTabular}
\caption{Execution accuracy by schema linker on Spider}
\label{tab:e2e}
\end{table}


\begin{table}[ht]
\centering
\begin{tabular}{l|cccc}
\hline
Linker  & {\small Simple}  & {\small Moderate}  & {\small Challenging}  & Total \\
\hline
GT$_{t}$  & 72.54 & 55.27 & 44.44 & 64.67 \\
GT$_{c}$  & 81.62 & 63.66 & 52.08 & 73.40 \\
GT$_{f}$  & 85.30 & 68.82 & 56.94 & 77.64 \\
\hline
Gen     & 65.30 & 52.69 & 46.53 & 59.71 \\
ExSL$_{c}$ & 68.54 & 53.76 & 45.14 & 61.86 \\
ExSL$_{f}$ & 70.05 & 55.70 & 43.75 & \textbf{63.23} \\
\hline
\end{tabular}
\caption{Execution accuracy by schema linker on BIRD}
\label{tab:e2e_bird}
\end{table}

\subsection{Precision/Recall Trade-Off}\label{sec:pr_tradeoff}

In this study, we explore how the $F_{\beta}$ scores of the schema linker influence the performance of SQL generation that utilizes these schema links. We analyze $F_{\beta}$ scores at various logit thresholds ranging from 0 to -6. For each threshold, we first calculate the $F_{\beta}$ score for the schema linker, then generate SQL queries based on the schema links, and finally assess the accuracy of these queries. Our goal is to determine the Spearman rank correlation between the $F_{\beta}$ scores, which serve as a component metric for schema linking, and the ultimate metric of SQL execution accuracy.

This analysis allows us to identify the $F_{\beta}$ score, particularly the $F_6$ score, as a critical component metric for evaluating schema linkers. We observe that selecting a logit threshold based on the $F_6$ score consistently yields the best or second-best results in terms of SQL generation accuracy across all considered thresholds.

The $F_{\beta}$ score is calculated as follows:

\begin{equation}\label{eq:fscore}
F_{\beta} = (1 + \beta^2) \cdot \frac{\text{precision} \cdot \text{recall}}{(\beta^2 \cdot \text{precision}) + \text{recall}}
\end{equation}

This formula integrates both precision and recall of the schema linker, adjusting their importance through the $\beta$ parameter. A higher $\beta$ value places more emphasis on recall, making it crucial in scenarios where missing relevant links is particularly costly.

Table \ref{tab:correl} shows the correlation between the $F_{\beta}$ scores at different $\beta$ levels and the accuracy of SQL query execution over different variations of Spider. 
The table highlights recall weighted F-scores as effective metrics for schema linking, while selecting a precision / recall trade-off by $F_1$ score is negatively correlated with end-to-end execution accuracy over the considered logit thresholds.

\begin{table}[!htb]
\centering
\begin{NiceTabular}{l|rrrrr}
\hline
$\beta$ & Dev & DK & Syn & Real & Avg. \\
\hline
1  & -0.487 & -0.536 & -0.613 & -0.857 & -0.623 \\
2  & -0.108 & 0.214 & 0.162 & -0.143 & 0.031 \\
3  & 0.649 & 0.643 & 0.595 & 0.607 & 0.623 \\
4  & 0.847 & 0.821 & 0.847 & 0.893 & 0.852 \\
5  & 0.919 & 0.821 & 0.883 & 0.964 & 0.897 \\
6  & 0.991 & 0.714 & 0.937 & 1.000 & \textbf{0.911} \\
7  & 0.937 & 0.536 & 0.847 & 0.964 & 0.821 \\
\hline
\end{NiceTabular}
\caption{Correlation of $F_{\beta}$ schema linking score with SQL generation execution accuracy on Spider}
\label{tab:correl}
\end{table}

\begin{figure}
    \centering
    \includegraphics[width=0.9\linewidth]{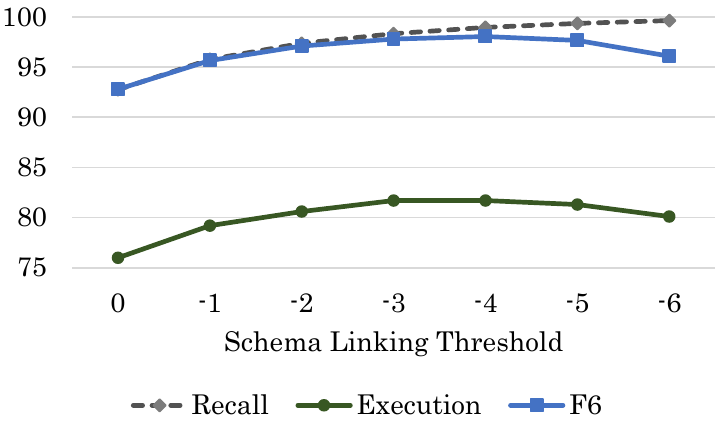}
    \caption{Sensitivity of SQL Generation performance on Spider Dev to schema linking threshold}
    \label{fig:threshold_sensitivity}
\end{figure}

Since we select a threshold for the schema linker, we examine how sensitive the performance of the SQL generation is to the selection of the threshold. Figure \ref{fig:threshold_sensitivity} shows that the threshold is not a sensitive hyperparameter, with similar performance in the region of -3 to -4. This graph also visually shows the $F_6$ score is strongly correlated with impact in end-to-end performance.

\subsection{Additional Analysis}
\label{sec:additional_analysis}

Another way to consider the impact of schema linking on the end-to-end task of SQL generation is to examine how the correctness of the generated SQL varies with the correctness of the schema linking.
Since both the schema linking and SQL generation are likely to be more accurate on easier Text-to-SQL instances, we also break this analysis down by question difficulty, grouping ``hard'' and ``extra-hard'' into just ``hard''.
Table \ref{tab:instance} buckets the instances in each benchmark by their schema linking F-score and gives the execution accuracy for each bucket.  We see that when the schema linking is more accurate we get a much higher execution accuracy.

Table \ref{tab:imperfect_recall} shows the performance of SQL generation on instances where schema linking did not identify all relevant columns.  The execution accuracy of such instances is very low, between 15 and 40 percent depending on the dataset.  The correctly executing queries are sometimes a result of the LLM's world knowledge, for example knowing the airport code for a city can eliminate the need to join on a table mapping airport codes to city names.  Rarely, SQL generation can correctly guess a column name.  Finally, many of the ``correctly executing'' queries in the subset of incomplete recall for schema linking give a result that is equivalent to the correct query's result by chance.

\begin{table}
\centering
\begin{NiceTabular}{r|rrrr}
\hline
& & \multicolumn{3}{c}{Schema Linking F-Score} \\
Dataset & Type & $<80\%$ & $\geq 80\%$ & $\geq 90\%$ \\
\hline
Dev     & All    & 76.52 & 82.12 & 88.17 \\  
        & \small{Easy}   & 87.29 & 92.31 & 96.77 \\
        & \small{Medium} & 73.39 & 86.34 & 90.70 \\
        & \small{Hard}   & 67.96 & 70.76 & 78.23 \\
DK      & All    & 63.41 & 72.51 & 77.44 \\
        & \small{Easy}   & 92.50 & 78.57 & 78.69 \\
        & \small{Medium} & 53.19 & 76.77 & 83.05 \\
        & \small{Hard}   & 44.44 & 63.64 & 68.97 \\
Syn     & All    & 66.08 & 77.21 & 82.76 \\  
        & \small{Easy}   & 74.09 & 76.36 & 91.30 \\
        & \small{Medium} & 69.57 & 82.27 & 86.49 \\
        & \small{Hard}   & 55.65 & 70.41 & 70.37 \\
Real    & All    & 58.54 & 80.52 & 85.12 \\  
        & \small{Easy}   & 79.41 & 90.67 & 94.44 \\
        & \small{Medium} & 58.70 & 90.45 & 95.70 \\
        & \small{Hard}   & 41.86 & 65.36 & 69.47 \\
\hline
\end{NiceTabular}
\caption{Execution accuracy on Spider bucketed by schema linking F-score.}
\label{tab:instance}
\end{table}

\begin{table}
\centering
\begin{NiceTabular}{l|rr}
\hline
Dataset & <100\% Recall & Execution   \\
\hline
Dev    & 24 (2.3\%) & 25.0 \\
DK     & 46 (8.6\%) & 34.8 \\
Syn    & 55 (5.3\%) & 16.4 \\
Real   & 41 (8.1\%) & 39.0 \\
\hline
\end{NiceTabular}
\caption{Performance of SQL generation on Spider without full recall in schema linking}
\label{tab:imperfect_recall}
\end{table}

\section{Conclusion}


By focusing on the relevant parts of the database schema, schema linking reduces the computational load and improves the accuracy of SQL query generation.  We introduce a novel method for extractive schema linking in Text-to-SQL systems, utilizing modern decoder-only large language models (LLMs). This approach not only demonstrates improved computational efficiency but also achieves superior accuracy compared to contemporary generative methods.   
Extensive evaluations on Spider, multiple variants, and BIRD establish a new state-of-the-art in schema linking.
The introduction of a probability-based model for schema linking allows for a controllable balance between precision and recall.
SQL generation also benefits from the fine-grained schema linking we introduce, which predicts the role or roles of each relevant column. Furthermore, we establish important links between metrics of schema linking performance at the component level and the end-to-end impact on SQL query generation.  The $F_6$ score, weighting recall much more heavily than precision, is well correlated with SQL generation accuracy.

\bibliography{anthology,custom}

\begin{thebibliography}{31}
\expandafter\ifx\csname natexlab\endcsname\relax\def\natexlab#1{#1}\fi

\bibitem[{Achiam et~al.(2023)Achiam, Adler, Agarwal, Ahmad, Akkaya, Aleman, Almeida, Altenschmidt, Altman, Anadkat et~al.}]{gpt-4}
Josh Achiam, Steven Adler, Sandhini Agarwal, Lama Ahmad, Ilge Akkaya, Florencia~Leoni Aleman, Diogo Almeida, Janko Altenschmidt, Sam Altman, Shyamal Anadkat, et~al. 2023.
\newblock Gpt-4 technical report.
\newblock \emph{arXiv preprint arXiv:2303.08774}.

\bibitem[{Deng et~al.(2021)Deng, Awadallah, Meek, Polozov, Sun, and Richardson}]{spider-realistic}
Xiang Deng, Ahmed~Hassan Awadallah, Christopher Meek, Oleksandr Polozov, Huan Sun, and Matthew Richardson. 2021.
\newblock \href {https://doi.org/10.18653/v1/2021.naacl-main.105} {Structure-grounded pretraining for text-to-{SQL}}.
\newblock In \emph{Proceedings of the 2021 Conference of the North American Chapter of the Association for Computational Linguistics: Human Language Technologies}, pages 1337--1350, Online. Association for Computational Linguistics.

\bibitem[{Dong et~al.(2023)Dong, Zhang, Ge, Mao, Gao, Lin, Lou et~al.}]{c3}
Xuemei Dong, Chao Zhang, Yuhang Ge, Yuren Mao, Yunjun Gao, Jinshu Lin, Dongfang Lou, et~al. 2023.
\newblock C3: Zero-shot text-to-sql with chatgpt.
\newblock \emph{arXiv preprint arXiv:2307.07306}.

\bibitem[{Gan et~al.(2021{\natexlab{a}})Gan, Chen, Huang, Purver, Woodward, Xie, and Huang}]{spider-syn}
Yujian Gan, Xinyun Chen, Qiuping Huang, Matthew Purver, John~R. Woodward, Jinxia Xie, and Pengsheng Huang. 2021{\natexlab{a}}.
\newblock \href {https://doi.org/10.18653/v1/2021.acl-long.195} {Towards robustness of text-to-{SQL} models against synonym substitution}.
\newblock In \emph{Proceedings of the 59th Annual Meeting of the Association for Computational Linguistics and the 11th International Joint Conference on Natural Language Processing (Volume 1: Long Papers)}, pages 2505--2515, Online. Association for Computational Linguistics.

\bibitem[{Gan et~al.(2021{\natexlab{b}})Gan, Chen, and Purver}]{spider-dk}
Yujian Gan, Xinyun Chen, and Matthew Purver. 2021{\natexlab{b}}.
\newblock \href {https://doi.org/10.18653/v1/2021.emnlp-main.702} {Exploring underexplored limitations of cross-domain text-to-{SQL} generalization}.
\newblock In \emph{Proceedings of the 2021 Conference on Empirical Methods in Natural Language Processing}, pages 8926--8931, Online and Punta Cana, Dominican Republic. Association for Computational Linguistics.

\bibitem[{Gao et~al.(2023)Gao, Wang, Li, Sun, Qian, Ding, and Zhou}]{dail-sql}
Dawei Gao, Haibin Wang, Yaliang Li, Xiuyu Sun, Yichen Qian, Bolin Ding, and Jingren Zhou. 2023.
\newblock Text-to-sql empowered by large language models: A benchmark evaluation.
\newblock \emph{arXiv preprint arXiv:2308.15363}.

\bibitem[{Guo et~al.(2024)Guo, Zhu, Yang, Xie, Dong, Zhang, Chen, Bi, Wu, Li, Luo, Xiong, and Liang}]{deepseek-coder}
Daya Guo, Qihao Zhu, Dejian Yang, Zhenda Xie, Kai Dong, Wentao Zhang, Guanting Chen, Xiao Bi, Y.~Wu, Y.~K. Li, Fuli Luo, Yingfei Xiong, and Wenfeng Liang. 2024.
\newblock Deepseek-coder: When the large language model meets programming -- the rise of code intelligence.
\newblock \emph{arXiv preprint arXiv:2401.14196}.

\bibitem[{Hochreiter and Schmidhuber(1997)}]{lstm}
Sepp Hochreiter and J{\"u}rgen Schmidhuber. 1997.
\newblock Long short-term memory.
\newblock \emph{Neural computation}, 9(8):1735--1780.

\bibitem[{Iyer et~al.(2017)Iyer, Konstas, Cheung, Krishnamurthy, and Zettlemoyer}]{iyer-etal-2017-learning}
Srinivasan Iyer, Ioannis Konstas, Alvin Cheung, Jayant Krishnamurthy, and Luke Zettlemoyer. 2017.
\newblock \href {https://doi.org/10.18653/v1/P17-1089} {Learning a neural semantic parser from user feedback}.
\newblock In \emph{Proceedings of the 55th Annual Meeting of the Association for Computational Linguistics (Volume 1: Long Papers)}, pages 963--973, Vancouver, Canada. Association for Computational Linguistics.

\bibitem[{Lei et~al.(2020{\natexlab{a}})Lei, Wang, Ma, Gan, Lu, Kan, and Chua}]{slsql}
Wenqiang Lei, Weixin Wang, Zhixin Ma, Tian Gan, Wei Lu, Min-Yen Kan, and Tat-Seng Chua. 2020{\natexlab{a}}.
\newblock \href {https://doi.org/10.18653/v1/2020.emnlp-main.564} {Re-examining the role of schema linking in text-to-{SQL}}.
\newblock In \emph{Proceedings of the 2020 Conference on Empirical Methods in Natural Language Processing (EMNLP)}, pages 6943--6954, Online. Association for Computational Linguistics.

\bibitem[{Lei et~al.(2020{\natexlab{b}})Lei, Wang, Ma, Gan, Lu, Kan, and Chua}]{lei-etal-2020-examining}
Wenqiang Lei, Weixin Wang, Zhixin Ma, Tian Gan, Wei Lu, Min-Yen Kan, and Tat-Seng Chua. 2020{\natexlab{b}}.
\newblock \href {https://doi.org/10.18653/v1/2020.emnlp-main.564} {Re-examining the role of schema linking in text-to-{SQL}}.
\newblock In \emph{Proceedings of the 2020 Conference on Empirical Methods in Natural Language Processing (EMNLP)}, pages 6943--6954, Online. Association for Computational Linguistics.

\bibitem[{Li et~al.(2023{\natexlab{a}})Li, Zhang, Li, and Chen}]{resdsql}
Haoyang Li, Jing Zhang, Cuiping Li, and Hong Chen. 2023{\natexlab{a}}.
\newblock Resdsql: decoupling schema linking and skeleton parsing for text-to-sql.
\newblock In \emph{Proceedings of the Thirty-Seventh AAAI Conference on Artificial Intelligence and Thirty-Fifth Conference on Innovative Applications of Artificial Intelligence and Thirteenth Symposium on Educational Advances in Artificial Intelligence}, pages 13067--13075.

\bibitem[{Li et~al.(2024)Li, Hui, Qu, Yang, Li, Li, Wang, Qin, Geng, Huo et~al.}]{li2024can}
Jinyang Li, Binyuan Hui, Ge~Qu, Jiaxi Yang, Binhua Li, Bowen Li, Bailin Wang, Bowen Qin, Ruiying Geng, Nan Huo, et~al. 2024.
\newblock Can llm already serve as a database interface? a big bench for large-scale database grounded text-to-sqls.
\newblock \emph{Advances in Neural Information Processing Systems}, 36.

\bibitem[{Li et~al.(2023{\natexlab{b}})Li, Allal, Zi, Muennighoff, Kocetkov, Mou, Marone, Akiki, Li, Chim et~al.}]{starcoder}
Raymond Li, Loubna~Ben Allal, Yangtian Zi, Niklas Muennighoff, Denis Kocetkov, Chenghao Mou, Marc Marone, Christopher Akiki, Jia Li, Jenny Chim, et~al. 2023{\natexlab{b}}.
\newblock Starcoder: may the source be with you!
\newblock \emph{arXiv preprint arXiv:2305.06161}.

\bibitem[{Liu et~al.(2021)Liu, Yang, Zhang, Guo, Zhou, and Lou}]{eta-awakening}
Qian Liu, Dejian Yang, Jiahui Zhang, Jiaqi Guo, Bin Zhou, and Jian-Guang Lou. 2021.
\newblock \href {https://doi.org/10.18653/v1/2021.findings-acl.100} {Awakening latent grounding from pretrained language models for semantic parsing}.
\newblock In \emph{Findings of the Association for Computational Linguistics: ACL-IJCNLP 2021}, pages 1174--1189, Online. Association for Computational Linguistics.

\bibitem[{Liu et~al.(2019)Liu, Ott, Goyal, Du, Joshi, Chen, Levy, Lewis, Zettlemoyer, and Stoyanov}]{roberta}
Yinhan Liu, Myle Ott, Naman Goyal, Jingfei Du, Mandar Joshi, Danqi Chen, Omer Levy, Mike Lewis, Luke Zettlemoyer, and Veselin Stoyanov. 2019.
\newblock Roberta: A robustly optimized bert pretraining approach.
\newblock \emph{arXiv preprint arXiv:1907.11692}.

\bibitem[{Luo et~al.(2023)Luo, Xu, Zhao, Sun, Geng, Hu, Tao, Ma, Lin, and Jiang}]{wizardcoder}
Ziyang Luo, Can Xu, Pu~Zhao, Qingfeng Sun, Xiubo Geng, Wenxiang Hu, Chongyang Tao, Jing Ma, Qingwei Lin, and Daxin Jiang. 2023.
\newblock Wizardcoder: Empowering code large language models with evol-instruct.
\newblock \emph{arXiv preprint arXiv:2306.08568}.

\bibitem[{Pennington et~al.(2014)Pennington, Socher, and Manning}]{pennington-etal-2014-glove}
Jeffrey Pennington, Richard Socher, and Christopher Manning. 2014.
\newblock \href {https://doi.org/10.3115/v1/D14-1162} {{G}lo{V}e: Global vectors for word representation}.
\newblock In \emph{Proceedings of the 2014 Conference on Empirical Methods in Natural Language Processing ({EMNLP})}, pages 1532--1543, Doha, Qatar. Association for Computational Linguistics.

\bibitem[{Pourreza and Rafiei(2023)}]{din-sql}
Mohammadreza Pourreza and Davood Rafiei. 2023.
\newblock Din-sql: Decomposed in-context learning of text-to-sql with self-correction.
\newblock \emph{arXiv preprint arXiv:2304.11015}.

\bibitem[{Pourreza and Rafiei(2024)}]{dts-sql}
Mohammadreza Pourreza and Davood Rafiei. 2024.
\newblock Dts-sql: Decomposed text-to-sql with small large language models.
\newblock \emph{arXiv preprint arXiv:2402.01117}.

\bibitem[{Rai et~al.(2023)Rai, Wang, Zhou, and Yao}]{rai-etal-2023-improving}
Daking Rai, Bailin Wang, Yilun Zhou, and Ziyu Yao. 2023.
\newblock \href {https://doi.org/10.18653/v1/2023.acl-short.15} {Improving generalization in language model-based text-to-{SQL} semantic parsing: Two simple semantic boundary-based techniques}.
\newblock In \emph{Proceedings of the 61st Annual Meeting of the Association for Computational Linguistics (Volume 2: Short Papers)}, pages 150--160, Toronto, Canada. Association for Computational Linguistics.

\bibitem[{Scholak et~al.(2021)Scholak, Schucher, and Bahdanau}]{scholak-etal-2021-picard}
Torsten Scholak, Nathan Schucher, and Dzmitry Bahdanau. 2021.
\newblock \href {https://doi.org/10.18653/v1/2021.emnlp-main.779} {{PICARD}: Parsing incrementally for constrained auto-regressive decoding from language models}.
\newblock In \emph{Proceedings of the 2021 Conference on Empirical Methods in Natural Language Processing}, pages 9895--9901, Online and Punta Cana, Dominican Republic. Association for Computational Linguistics.

\bibitem[{Shaw et~al.(2021)Shaw, Chang, Pasupat, and Toutanova}]{shaw-etal-2021-compositional}
Peter Shaw, Ming-Wei Chang, Panupong Pasupat, and Kristina Toutanova. 2021.
\newblock \href {https://doi.org/10.18653/v1/2021.acl-long.75} {Compositional generalization and natural language variation: Can a semantic parsing approach handle both?}
\newblock In \emph{Proceedings of the 59th Annual Meeting of the Association for Computational Linguistics and the 11th International Joint Conference on Natural Language Processing (Volume 1: Long Papers)}, pages 922--938, Online. Association for Computational Linguistics.

\bibitem[{Suhr et~al.(2020)Suhr, Chang, Shaw, and Lee}]{suhr-etal-2020-exploring}
Alane Suhr, Ming-Wei Chang, Peter Shaw, and Kenton Lee. 2020.
\newblock \href {https://doi.org/10.18653/v1/2020.acl-main.742} {Exploring unexplored generalization challenges for cross-database semantic parsing}.
\newblock In \emph{Proceedings of the 58th Annual Meeting of the Association for Computational Linguistics}, pages 8372--8388, Online. Association for Computational Linguistics.

\bibitem[{Vaswani et~al.(2017)Vaswani, Shazeer, Parmar, Uszkoreit, Jones, Gomez, Kaiser, and Polosukhin}]{attention}
Ashish Vaswani, Noam Shazeer, Niki Parmar, Jakob Uszkoreit, Llion Jones, Aidan~N Gomez, {\L}ukasz Kaiser, and Illia Polosukhin. 2017.
\newblock Attention is all you need.
\newblock \emph{Advances in neural information processing systems}, 30.

\bibitem[{Wang et~al.(2020)Wang, Shin, Liu, Polozov, and Richardson}]{wang-etal-2020-rat}
Bailin Wang, Richard Shin, Xiaodong Liu, Oleksandr Polozov, and Matthew Richardson. 2020.
\newblock \href {https://doi.org/10.18653/v1/2020.acl-main.677} {{RAT-SQL}: Relation-aware schema encoding and linking for text-to-{SQL} parsers}.
\newblock In \emph{Proceedings of the 58th Annual Meeting of the Association for Computational Linguistics}, pages 7567--7578, Online. Association for Computational Linguistics.

\bibitem[{Wang et~al.(2021)Wang, Wang, Joty, and Hoi}]{wang-etal-2021-codet5}
Yue Wang, Weishi Wang, Shafiq Joty, and Steven~C.H. Hoi. 2021.
\newblock \href {https://doi.org/10.18653/v1/2021.emnlp-main.685} {{C}ode{T}5: Identifier-aware unified pre-trained encoder-decoder models for code understanding and generation}.
\newblock In \emph{Proceedings of the 2021 Conference on Empirical Methods in Natural Language Processing}, pages 8696--8708, Online and Punta Cana, Dominican Republic. Association for Computational Linguistics.

\bibitem[{Yu et~al.(2018)Yu, Zhang, Yang, Yasunaga, Wang, Li, Ma, Li, Yao, Roman, Zhang, and Radev}]{yu-etal-2018-spider}
Tao Yu, Rui Zhang, Kai Yang, Michihiro Yasunaga, Dongxu Wang, Zifan Li, James Ma, Irene Li, Qingning Yao, Shanelle Roman, Zilin Zhang, and Dragomir Radev. 2018.
\newblock \href {https://doi.org/10.18653/v1/D18-1425} {{S}pider: A large-scale human-labeled dataset for complex and cross-domain semantic parsing and text-to-{SQL} task}.
\newblock In \emph{Proceedings of the 2018 Conference on Empirical Methods in Natural Language Processing}, pages 3911--3921, Brussels, Belgium. Association for Computational Linguistics.

\bibitem[{Zelle and Mooney(1996)}]{geoquery}
John~M Zelle and Raymond~J Mooney. 1996.
\newblock Learning to parse database queries using inductive logic programming.
\newblock In \emph{Proceedings of the national conference on artificial intelligence}, pages 1050--1055.

\bibitem[{Zhang et~al.(2023)Zhang, Cao, Chen, Xu, and Yu}]{act-sql}
Hanchong Zhang, Ruisheng Cao, Lu~Chen, Hongshen Xu, and Kai Yu. 2023.
\newblock \href {https://doi.org/10.18653/v1/2023.findings-emnlp.227} {{ACT}-{SQL}: In-context learning for text-to-{SQL} with automatically-generated chain-of-thought}.
\newblock In \emph{Findings of the Association for Computational Linguistics: EMNLP 2023}, pages 3501--3532, Singapore. Association for Computational Linguistics.

\bibitem[{Zhong et~al.(2017)Zhong, Xiong, and Socher}]{wikisql}
Victor Zhong, Caiming Xiong, and Richard Socher. 2017.
\newblock Seq2sql: Generating structured queries from natural language using reinforcement learning.
\newblock \emph{arXiv preprint arXiv:1709.00103}.

\end{thebibliography}

\appendix

\newpage


\begin{figure*}
    \centering
    \includegraphics[width=0.9\linewidth]{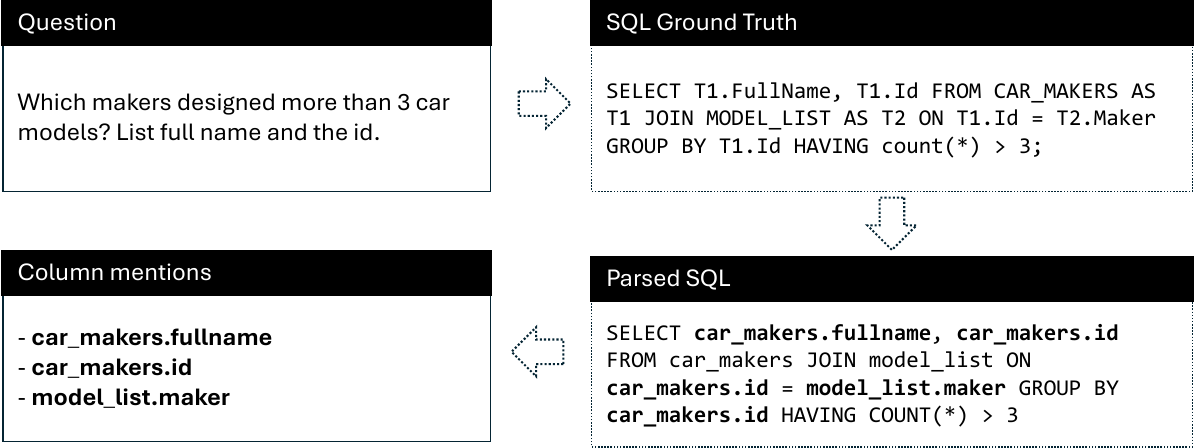}
    \caption{Ground truth generation for schema linking}
    \label{fig:ground_truth2}
\end{figure*}

\begin{table*}[!htb]
    \begin{minipage}[t]{0.48\linewidth}
    \section{Schema Linking Ground Truth}
\label{sec:sl-sql-gt}
        Figure \ref{fig:ground_truth2} shows an example of our process for constructing schema linking ground truth. To find the ground truth schema links for a question in the training set, we start from the ground truth SQL query. We parse the query and identify all column mentions. We will also release this ground truth alongside our open source code.
        
        The adjacent question and SQL pairs show examples of SL-SQL schema linking annotations. In the first case the schema linking annotation correctly requires the \texttt{employee.lastname}, which is not in the ground truth SQL query where \texttt{customer.lastname} is selected instead.   
        Question 2, in contrast, shows the mixture of automatic and manual annotation can add unneeded tables and columns (\texttt{customers} is not needed). 
        
        The most frequent cause of mismatch between SL-SQL annotations and our ground truth is when the SL-SQL annotation omits a column because it does not have a close string match with the question, and annotators neglected to manually add it.
\end{minipage}\hfill
    \begin{minipage}[t]{0.48\linewidth}

    \centering
    \small
    \begin{tcolorbox}[colback=gray!10,colframe=black,title=Question 1]
    What are the last names of employees who serve at most 20 customers?
    \end{tcolorbox}
    
    \begin{tcolorbox}[colback=gray!10,colframe=black,title=SQL Query 1]
    \begin{lstlisting}[basicstyle=\ttfamily,columns=fullflexible,aboveskip=-5pt,belowskip=-5pt]
    SELECT T1.LastName
      FROM CUSTOMER AS T1
      JOIN EMPLOYEE AS T2 
      ON T1.SupportRepId = T2.EmployeeId
     GROUP BY T1.SupportRepId
    HAVING COUNT(*) <= 20
    \end{lstlisting}
    \end{tcolorbox}
    
    \begin{tabular}{|c|c|}
    \hline
    \textbf{Tables} & \textbf{Columns} \\ \hline
    employee & employee.lastname \\ \hline
    customer &  \\ \hline
    \end{tabular}

    \centering
    \small
    \begin{tcolorbox}[colback=gray!10,colframe=black,title=Question 2]
    Show all customer ids and the number of accounts for each customer.
    \end{tcolorbox}
    
    \begin{tcolorbox}[colback=gray!10,colframe=black,title=SQL Query 2]
    \begin{lstlisting}[basicstyle=\ttfamily,columns=fullflexible,aboveskip=-5pt,belowskip=-5pt]
    SELECT customer_id, COUNT(*)
      FROM accounts 
     GROUP BY customer_id
    \end{lstlisting}
    \end{tcolorbox}
    
    \begin{tabular}{|c|c|}
    \hline
    \textbf{Tables} & \textbf{Columns} \\ \hline
    accounts & accounts.customer\_id \\ \hline
    customers &  \\ \hline
    \end{tabular}

\end{minipage}
\end{table*}

\FloatBarrier

\section{Open Source Model Comparison}
\label{sec:open-source-model-comp}
For our initial experiments we attempted both schema linking and SQL generation with a number of open source models.  We used StarCoder \cite{starcoder}, WizardCoder \cite{wizardcoder} and DeepSeek Coder \cite{deepseek-coder}. These models were chosen for their permissive licenses, competitive performance and for the availability of models with 15B parameters or less - to permit many experiments with a reasonable computational budget. We find that for both schema linking and SQL generation the DeepSeek Coder model provides the best performance while also using the fewest parameters. 

Table \ref{tab:os_models_schema_link} gives the initial performance of these models at schema linking.  
ROC refers to the area under the Receiver Operating Curve, which is the probability that a random relevant schema item is ranked higher than a random irrelevant schema item.  This is the metric reported by RESDSQL for their cross-encoder schema linker.  
PR refers to the area under the precision/recall curve, also known as average precision.  $F_6$ is the F score weighted to value recall at six times the weight of precision.  We justify this metric in section \ref{sec:pr_tradeoff}.

Table \ref{tab:os_models_sql_gen} gives the performance of the models for SQL generation using  basic generative schema linking of Section \ref{gensl}.  

\begin{table}
\centering
\begin{NiceTabular}{l|rrrr}
\hline
Model & Param & ROC & PR & $F_6$  \\
\hline
StarCoder      & 15B  & 99.52 & 96.92 & 97.79 \\
WizardCoder    & 15B  & 99.53 & 96.60 & 97.78 \\
DeepSeek       & 6.7B & \textbf{99.64} & \textbf{97.68} & \textbf{98.04} \\
\hline
\end{NiceTabular}
\caption{Base performance for open source models on Spider schema linking}
\label{tab:os_models_schema_link}
\end{table}

\begin{table}
\centering
\begin{NiceTabular}{l|rr}
\hline
Model & Parameters & Execution   \\
\hline
StarCoder      & 15B  & 78.1 \\
WizardCoder    & 15B  & 76.2 \\
DeepSeek       & 6.7B & \textbf{80.9} \\
\hline
\end{NiceTabular}
\caption{Base performance for open source models on Spider SQL generation}
\label{tab:os_models_sql_gen}
\end{table}

\section{Hyperparameters}
\label{sec:hyperparameters}

We use the same training hyperparameters for generative, ExSL$_c$, and ExSL$_f$.

\begin{table}[h!]
\centering
\begin{NiceTabular}{r|r}
\hline
Hyperparameter & Value   \\
\hline
Base Model & DeepSeek Coder 6.7B \\
Max Input Tokens & 3000 \\
Optimizer & AdamW \\
LR & 5e-6 \\
Weight Decay & 0.0 \\
Batch Size & 16 \\
Train Epochs & 2 \\
GPU & A100 80 GB \\
Time & 9-10 hours \\
\hline
\end{NiceTabular}
\caption{Hyperparameters for training schema linking}
\label{tab:hypers}
\end{table}

\end{document}